\documentclass[11pt]{article}
\usepackage{hyperref,amsmath,natbib}
\pdfoutput=1

\begin{document}
\title{Stirring faces: mixing in a quiescent fluid}
\author{Steven L. Brunton and Clarence W. Rowley \\
\date{}
\\\vspace{6pt}
Mechanical and Aerospace Engineering,\\
Princeton University, Princeton, NJ 08540, USA}
\maketitle
%% The abstract (in this file, and that submitted as text to arXiv) should
%include the exact phrase
%% "fluid dynamics video" or "fluid dynamics videos"
\begin{abstract}
This fluid dynamics video depicts the mixing that occurs as a two-dimensional flat plate plunges sinusoidally in a quiescent fluid.  Finite-time Lyapunov exponents reveal sets that are attracting or repelling.  As the flow field develops, strange faces emerge.
\end{abstract}
% main text

This video depicts the fluid mixing that occurs as a two-dimensional flat plate plunges sinusoidally in a quiescent fluid.  The flow is simulated using the immersed boundary projection method of~\cite{taira:07ibfs,taira:fastIBPM}.  Length is nondimensionalized by the chord length, and velocity is nondimensionalized so that a unit free-stream velocity would correspond to a Reynolds number of $\text{Re}=100$.  The plunging motion is sinusoidal with prescribed vertical motion: $y(t) = A\sin(2\pi f t)$, with $A=0.5$ and $f=0.5$ in dimensionless units.  

Separated flow structures are visualized by plotting fields of the finite-time Lyapunov exponent (FTLE)~\citep{ftle_02:Ha,shadden:07,brunton:fFTLE}.  These structures represent material lines that either attract or repel particles in forward time.  The duration of the particle integration used to compute these FTLE fields is $T=4$ (dimensionless time).  

As the unsteady flow develops, strange faces begin to appear in the FTLE field.  Although this is an illusion, brought about by symmetry in the vortex cores, the variety of expressions is striking.

%\item In the Abstract (in the LaTeX file and in the text submitted
%to arXiv), the exact phrase ``fluid dynamics video" or ``fluid
%dynamics videos".  This is to facilitate subsequent searching.

%
%\begin{spacing}{1.}
%\small{
%%\setlength{\bibsep}{1pt}
%\bibliographystyle{Science}
%\bibliography{/Users/sbrunton/Dropbox/PAPERS/master}}
%\end{spacing}
\setlength{\bibsep}{1pt}

\end{document}